\documentstyle[preprint,aps]{revtex}
\begin{document}
\draft
\title{\bf Planetoid Strings : Solutions and Perturbations}
\author{Sayan Kar \thanks{Electronic Address :
sayan@iucaa.ernet.in}} 
\address{Inter--University Centre for Astronomy and Astrophysics,\\
Post Bag 4, Ganeshkhind, Pune, 411 007, INDIA}
\author{Swapna Mahapatra \thanks{Electronic Address : 
swapna@qft2.physik.hu-berlin.de}}
\address{Institut f\"ur Physik, Humboldt Universit\"at zu 
Berlin, \\
Invalidenstr. 110, D-10115 Berlin, 
GERMANY}
\maketitle
\parshape=1 0.75in 5.5in
\begin{abstract}
A novel ansatz for solving the string equations of motion
and constraints in generic curved backgrounds, 
namely the {\em planetoid ansatz}, was proposed recently by
some authors. We construct several specific
examples of planetoid strings in curved backgrounds
which include Lorentzian wormholes, spherical Rindler spacetime
and the 2+1 dimensional black hole. A semiclassical quantisation 
is performed and the Regge relations for the planetoids are obtained.
The general equations for the study of small perturbations about 
these solutions are written down using the standard, manifestly covariant 
formalism. Applications to special cases such as those of
planetoid strings in Minkowski and spherical Rindler spacetimes
are also presented.
\end{abstract}

PACS number(s) : 11.25.-w, 04.70.-s, 98.80.Cq
\pacs{}

\newpage
\section{\bf Introduction}

In the context of cosmic as well as fundamental strings,
the analysis of the classical string equations of motion and 
constraints in generic curved backgrounds {\cite{dvs:87}}
has become an
active area of research over the last decade or so (for a recent
review and references see {\cite{s:sqg}})
Solutions representing string configurations,
which essentially correspond to timelike embedded minimal surfaces, 
are difficult to obtain largely
due to the nonlinear and coupled nature of the relevant equations.
Therefore,
the attitude has been to proceed by proposing a generic ansatz
based on symmetries or simplifying assumptions
which reduce the complicated set of equations to a tractable form.
Among the various proposals till date, we have the 
stationary string ansatz {\cite{stat:ref}}, dynamic circular strings
{\cite{dyn:ref}}
and more recently, planetoid string configurations {\cite{dve:prd96}} as
well as rigidly rotating strings {\cite{fhdv:hepth}}. It is
worth mentioning that the planetoid and rigidly rotating
strings are both special cases of an ansatz proposed
earlier by Larsen and Sanchez {\cite{ls:ans}}.  Target spaces with
metrics such as the Schwarzschild, Kerr-Newman, Robertson--Walker,
cosmic strings, wormholes etc. have been chosen and explicit
string configurations obtained in these backgrounds.
Once specific configurations are known, the obvious
next question that emerges is about their stability. This
turns out to be related to the second variation of the
action (Nambu--Goto or its generalisations)
and the corresponding Jacobi equations {\cite{glfc:ref}}. 
Perturbative stability
depends crucially on the analysis of these equations.
String propagation in an exact, stringy four-dimensional black hole 
background and 
the perturbations about extremal configurations
have also been studied recently {\cite{string:swap}}.
Furthermore, nonperturbative effects which include the
formation of cusps and kinks on the world surface of the
string is governed by the character of solutions of the
generalised Raychaudhuri equations {\cite{cgk:prd96}}.

This paper deals with planetoid strings. First we obtain
specific solutions in certain well known backgrounds. Thereafter,
we discuss small perturbations
about these configurations. The backgrounds chosen include
the Ellis geometry (a Lorentzian wormhole), the spherical Rindler 
spacetime, the Minkowski spacetime and the 2+1 dimensional
BTZ black hole {\cite{btz:2+1}}. We also consider semi-classical quantization 
of strings 
in these backgrounds and compute physical quantities such as the classical 
action, mass, reduced action and angular momentum.
The quantisation condition for each case is written explicitly.

Notations and sign conventions in the paper follow the 
norms of Misner, Thorne and Wheeler {\cite{mtw:book}}. 

\section{\bf Planetoid Strings: Formalism}

We first briefly discuss general planetoid strings, quote the 
ansatz and the resulting equations which we solve for specific
backgrounds later. 

The generic background metric (taking a $\theta = \frac{\pi}{2}$ section)
is taken to be of the form :

\begin{equation}
ds^{2} = g_{tt} dt^{2} + g_{rr}dr^{2} + 2g_{t\phi}dtd\phi + g_{\phi\phi}
d\phi^{2}
\end{equation}

The planetoid ansatz is given as{\cite{dve:prd96}},

\begin{equation}
t = t_{0} + \alpha \tau \quad ; \quad \phi = \phi_{0} + \beta \tau
\quad ; \quad r = r(\sigma)
\end{equation}
where, $\tau$ and $\sigma$ are the time-like and space-like coordinates
on the worldsheet respectively. $\alpha$ and $\beta$ are two arbitrary
constants. Assuming $\beta = 0$ would give us the usual stationary
strings. Note that the planetoid ansatz is a special case
of the one proposed by Larsen and Sanchez {\cite{ls:ans}} where
the constants $t_{0}$ and $\phi_{0}$ are replaced by
general functions $t_{0}(\sigma)$ and $\phi_{0} (\sigma)$
respectively. 
A word about the name `planetoid'. The ansatz above is a sort of
generalisation of the ansatz one would take if one deals with
the embedded curves along which planets move in their orbit.
Hence it is perhaps appropriate to call these kinds of worldsheets
`planetoids' -- a name which drives home the message that these
are related to planetary orbits while being surfaces as opposed to
curves.

From the bosonic string equations of motion and constraints
one arrives at the first order equation, which one needs to solve
in order to get a planetoid string. This is given as :

\begin{equation}
\left ( \frac{dr}{d\sigma} \right )^{2} = - g^{rr} \left [ \alpha^{2}
g_{tt} + 2\alpha\beta g_{t\phi} + \beta^{2} g_{\phi\phi}
\right ]
\end{equation}
where the right hand side can be identified with the negative of a potential 
$V(r)$.
However, it is more convenient to work with $\tilde V(r)$ which is defined as,

\begin{equation}
\tilde V(r) = \frac{V(r)}{\alpha^2} = g^{rr} \left [ g_{tt} + \frac{2\beta}
{\alpha} g_{t\phi} + \frac{\beta^2}{\alpha^2} g_{\phi\phi}\right ]
\end{equation}

The induced metric on the world-sheet of the string is given as :

\begin{equation}
ds_{I}^{2} = \left ( \frac{dr}{d\sigma} \right )^{2} g_{rr}
\left [ -d\tau^{2} + d\sigma^{2} \right ]
\end{equation}

By choosing a conformal gauge in which the induced metric is
diagonal and conformal to Minkowski spacetime in two dimensions,
we automatically satisfy the constraint equations
($g_{\mu\nu}{\dot x}^{\mu}{\dot x}^{\nu} + g_{\mu\nu}{x^{\prime}}^{\mu}
{x^{\prime}}^{\nu} = 0 ; \quad g_{\mu\nu}{\dot x}^{\mu}
{x^{\prime}}^{\nu}$
= 0), where dot and prime denote differentiations with respect to world-sheet
coordinates $\tau$ 
and $\sigma$ respectively and $\mu$, $\nu$ are space-time indices.
We confine ourselves largely to spherically symmetric, static backgrounds
for which the basic equation to solve turns out to be :

\begin{equation}
\frac{dr}{d\sigma} = \pm \sqrt{ \left (1-\frac{b(r)}{r} \right )
\left [ \alpha^{2}e^{2\psi (r)} - r^{2}\beta^{2} \right ]}
\end{equation}
where our background metric is now assumed as diagonal and for a 
$\theta = \frac{\pi}{2}$ section, it is given as :

\begin{equation}
ds^{2} = -e^{2\psi(r)}dt^{2} + \frac{dr^{2}}{1-\frac{b(r)}{r}}
+ r^{2}d\phi^{2}
\end{equation}

When is the induced metric on a planetoid string Minkowskian?
By looking at the expression for the induced metric one can
easily say that this happens if :

\begin{equation}
\left (\frac{dr}{d\sigma}\right )^{2} = C^{2}g^{rr}
\end{equation}

For spherically symmetric, static metrics this turns out to be
a very stringent constraint on the red--shift function $\psi(r)$,
which should satisfy,

\begin{equation}
e^{2\psi (r)} = \frac{C^{2} + r^{2}\beta ^{2}}{\alpha^{2}}
\end{equation}

Additionally, we observe that the existence of a zero in the
conformal factor in the induced metric would indicate the
existence of a singularity on the worldsheet. Specifically,
if $r=r_{0}$ is a zero of the expression for the conformal
factor we must have :

\begin{equation}
e^{2\psi(r_0)}\alpha^{2} = \beta^{2}r_{0}^{2}
\end{equation}

If $r_{0}$ coincides with the horizon $e^{2\psi (r_0)} = 0$
then we can only have $r_{0} = 0$. There maybe other points
in the geometry where this could be satisfied too regardless
of whether the geometry has a horizon or not. On the
other hand if $e^{2\psi} = 1$ (i.e. an ultrastatic metric)
we can clearly see that $r=\frac{\alpha}{\beta}$ is the
point where the worldsheet will become singular.  
These facts will be generic features of all the  solutions
to be discussed below.
 
Let us also consider planetoids in generic time--dependent
backgrounds of the form :

\begin{equation}
ds^{2} = \Omega ^{2} (t) \left ( -dt^{2} + \frac{dr^{2}}{1 - \frac{b(r)}{r}}
+ r^{2} d\Omega_{2}^{2}\right )
\end{equation}
 
It can be shown that there will be no planetoid solutions
in time-dependent backgrounds of the above type (which includes
the FRW models too). To see this let us look at the string
equation of motion for the coordinate $\phi$.
With the substitution of the planetoid ansatz, we find that
the equation reduces to the requirement :

\begin{equation}
-\alpha \beta \frac{\dot \Omega}{\Omega} = 0
\end{equation}

Since $\alpha$ or $\beta$ cannot be taken to be zero one
needs $\Omega (t)$ to be a constant. 

Also note that the planetoid ansatz is incompatible with null
(tensionless) strings as has been pointed out in {\cite{dl:hepth}}. 

We now move on towards solving the string equation of motion
 to obtain specific
planetoid string configuration in some well--known backgrounds.

\section{\bf Solutions in specific backgrounds}

{\sf (1) Spherically symmetric coordinate representation of Minkowski
spacetime}

In this case, the background metric (as given in the form in eqn. (7)) 
has $b(r) =0$ and $\psi (r) =0$. The planetoid solution is :
\begin{equation}
r = \frac{\alpha}{\beta} \vert \sin \beta \left (\sigma - \sigma_{0} \right )
\vert 
\end{equation}

The expression for $r(\sigma)$ is the modulus of the sine function. One
needs to consider the absolute value
 in order to define the worldsheet with proper
Neumann--type boundary conditions at the edge 
values $\sigma = \sigma_{0}\pm\frac{\pi}{2\beta}$. This, ofcourse results in a
{\em kink} in the metric and consequently a $\delta$--function
curvature singularity at $\sigma =\sigma_{0}$.  

The induced metric at all points $\sigma \neq 0$
 on the world-sheet is given as :

\begin{equation}
ds^{2} = \alpha^{2}\cos^{2} \beta (\sigma - \sigma_{0}) \left ( -d\tau^{2}
+ d\sigma^{2} \right )
\end{equation}
which is certainly not flat.

Also note that the metric on the worldsheet has a singularity at
$\sigma = \sigma_{0} \pm \frac{\pi}{2\beta}$. This can be
checked easily by calculating the Ricci scalar which turns out to
be :

\begin{equation}
^{2}R = \frac{2\beta^{2}}{\cos ^{4} \beta \left (\sigma - \sigma_{0}\right )}
\end{equation}

where we have excluded the $\delta$--function contribution at $\sigma =0$.
 
Therefore, it is necessary to restrict the worldsheet to the
domain $-\frac{\pi}{2\beta}<\sigma - \sigma_{0} < \frac{\pi}{2\beta}$.
   
(This planetoid solution is obtained in {\cite{dve:prd96}}, but the important point is
to realise is that the induced metric 
is not flat, is singular and therefore, as we shall see in the later sections 
of
this paper the
perturbation equations will be nontrivial by virtue of the
$K_{ab}^{i}K^{ab}_{i}$ term even though the Riemann tensor
term makes no contribution. ) 

{\sf (2) Spherical Rindler Spacetime}

Here $e^{2\psi (r)} = r^{2}$ and $ b(r) = 0$.
A word about the spherical Rindler spacetime. This is the generalisation
of the usual Rindler spacetime written in Cartesian coordinates
to a  spherisymmetric metric. The matter stress energy which
would be required to support such a geometry is however somewhat
strange. Firstly, spherical Rindler spacetime is not a flat
geometry--it has curvature. This can be checked by calculating
the Riemann tensor components. Moreover, the nonzero components
of $G_{\mu\nu}$  (in the frame--basis) are given as :

\begin{equation}
G_{00}= 0 \qquad ;\qquad G_{11}= \frac{2}{r^{2}} \qquad ;\qquad
G_{22}=G_{33}=\frac{1}{r^{2}}
\end{equation}

Defining the energy--momentum tensor $T_{\mu\nu}$ as given by the
$G_{\mu\nu}$ via Einstein's field equations, one can see that the
matter required to support spherical Rindler spacetime has an
equation of state $\tau=2p$ and also obeys the Weak Energy Condition 
(i.e. if $\rho$, $\tau$, $p$, $p$ are the diagonal components of the
$T_{\mu\nu}$ then one must have $\rho \ge 0, \rho +\tau \ge 0, rho+p\ge 0$ ).

The horizon of the spacetime ($r=0$) is also a singular point
in the sense of diverging Riemann curvature.  

The planetoid solution in this spacetime looks as follows:

\begin{equation}
r(\sigma) = \exp \left (\pm \sqrt {(\alpha^{2} - \beta^{2})} (\sigma +
 \sigma_{0})\right )
\end{equation}

with $\beta^{2} < \alpha^{2}$. The solution with the $+$ sign is
valid for $\sigma \le 0$ while the one with the $-$ sign
is for the domain $\sigma > 0$. These two solutions ofcourse match at
$\sigma = 0$ smoothly. The derivatives of $r(\sigma)$
however are not continuous at $\sigma = 0$. However, note that
the induced metric contains $\left (\frac{dr}{d\sigma}\right )^{2}$
and therefore the metric functions as well as
the extrinsic curvatures match smoothly across 
$\sigma = 0$. We need to have two
different solutions in the two domains of $\sigma$ in order
to satisfy the open string Neumann--type boundary conditions at
$\sigma = \pm \infty$. The string worldsheet here stretches from
$+\infty$ to $-\infty$ and describes a {\em folded} string with
the fold at $\sigma =0$. One can also adopt the
viewpoint that the above solution (say, the $+$ one) describes a 
semi--infinite string stretching from $r=0$ to infinity
with its only boundary at $r=0$ ($\sigma\rightarrow -\infty$). With $\beta^{2} > \alpha ^{2}$ we have
oscillatory solutions (they must be complex and therefore
are devoid of any physical relevance) 
and for $\beta^{2} = \alpha^{2}$, $r(\sigma)$ is
linear.
The worldsheet metric, however, is now flat. This is easily seen
by looking at the conformal factor in the induced metric which
goes as $e^{\pm \sqrt{\alpha^{2} -\beta^{2}} \sigma}$. For a
general 2D metric with a conformal factor $e^{2\rho}$ we have
$^{2}R = -2 e^{-2\rho}\partial_{+}\partial_{-}\rho$. Since $\rho$
is linear in this case we have $^{2}R = 0$ straightaway. 
However, recall the fact that spherical Rindler spacetime in
contrast to Rindler spacetime in Cartesian coordinates is 
a curved space-time and nonzero components of the Riemann tensor do
exist. Thus, spacetime curvature will contribute to the
perturbations, as we shall see later on. 

{\sf (3) Ellis geometry}

For this case, we have  $\psi(r) = 0$ and $b(r) = \frac{b_{0}^{2}}{r}$ in
the spherically symmetric metric.
This geometry represents a traversable Lorentzian wormhole
and was discussed first by Ellis {\cite{e:jmp71}} (see also
Morris and Thorne {\cite{mt:ajp88}}). The throat of the
wormhole (which corresponds also to the minimum value of $r$)
is at $r=b_{0}$. As is easily seen, the spacetime is asymptotically
flat--two asymptotic regions are connected by the wormhole tunnel. -

After explicit integration of the equation for $r(\sigma)$
we get the following expression for
the planetoid configuration :

\begin{equation}
r^{2} (\sigma) = \frac{1}{2} (A - B) \{ \frac {A + B}{A - B} \pm
\sin 2\beta (\sigma - \sigma_{0}) \}
\end{equation}
where $A = b_{0}^{2}, B = \frac{\alpha^{2}}{\beta^{2}}$.

If $A>B$ then $\frac{A+B}{A-B} > 1$ and both the $+$ and $-$ solutions
are valid. However, in this case $\frac{\alpha^{2}}{\beta^{2}} < r^{2} < 
b_{0}^{2}$.
But one needs $r \ge b_{0}$, otherwise the spacetime loses its Lorentzian
signature. Therefore, this solution is not possible physically.
On the other hand if $A<B$ then we have $b_{0}^{2} < r^{2} < \frac{
\alpha^{2}}{\beta^{2}}$ and the planetoid string extends from 
the minimum value of $r$ (i.e. $b_{0}$) upto the value $\frac{\alpha}
{\beta}$ which is perfectly allowed. The open string boundary
conditions are satisfied at the values $\sigma=\pm \frac{\pi}{4\beta}$
which correspond to $r$ values $b_{0}$ and $\frac{\alpha}{\beta}$.

The induced metric on the worldsheet is once again not flat and is given
(for the case $\frac{\alpha^{2}}{\beta^{2}} > b_{0}^{2}$) by :
 
\begin{equation}
ds_{I}^{2} = \frac{1}{2}\beta^{2} \left (\frac{\alpha^{2}}{\beta^{2}}
-b_{0}^{2} \right ) \left ( 1 \mp \sin 2\beta \sigma \right )
\left [ -d\tau^{2} + d\sigma^{2} \right ]
\end{equation}

The conformal factor becomes zero at the points 
$\sigma = \pm \frac{\pi}{4\beta}$
for the $-$ and $+$ solutions respectively. This corresponds to a
worldsheet singularity at those points. In spacetime, the worldsheet
singularity coincides with the location of the throat at $r=b_{0}$. 
 
{\sf (4) 2+1 dimensional black hole}

Here, we consider the $2 + 1$ dimensional black hole anti-de Sitter 
space-time obtained by Banados, Teitelboim and Zanelli {\cite{btz:2+1}}.
The metric is given by,

\begin{equation}
ds^2 = \left (M - \frac{r^2}{l^2} \right ) dt^2 + \frac{1}{\left (\frac{r^2}
{l^2} -
M + \frac{J^2}{4 r^2}\right )} dr^2 + r^2 d\phi^2 - J dt d\phi
\end{equation}
where, $M$ is the mass and $J$ is the angular momentum of the black hole.
For simplicity, we take the $M = 1$ and $J = 0$ limit of this metric. 
In this limit, there is only one horizon at $r = l$. The equation of motion 
obtained as a first order equation is given as, 

\begin{equation}
\left ( \frac{dr}{d\sigma}\right )^2 = \left (\frac{r^2}{l^2}
- 1 \right ) \left [\alpha^2 \left (\frac{r^2}{l^2} - 1 \right ) - 
\beta^2 r^2 \right ]
\end{equation}
which is solved in terms of incomplete elliptic integrals {\cite{gr:tab}}. 
The solution 
is given by,

\begin{equation}
\pm (\sigma - \sigma_0) = \frac{l}{\alpha} F(arcsin \left (\frac{r}{l}
\right ), k)
\end{equation}
where, $k^2 = 1 - \frac{\beta^2 l^2}{\alpha^2}$. Inverting the
above expression we obtain the form of $r(\sigma)$ which is given as
:
\begin{equation}
r(\sigma) = l \hspace{.05in}
  \vert sn \left [\pm \frac{ \alpha (\sigma - \sigma_{0})}{l}\right ] \vert 
\end{equation}
where $sn$ denotes the Jacobian elliptic function.
This string solution lies exclusively inside the horizon
of the black hole. However, as in the Minkowski spacetime solution,
one needs to take the absolute value in order to satisfy the
boundary conditions at the edges -- this results in a similar
{\em kink} in the metric and a $\delta$ function singularity at 
$\sigma =0$. The embedding function is plotted in Fig. 1. The edges of
the worldsheet are at those points where the derivative of the
function $r(\sigma)$ vanishes.  

The induced metric on the world-sheet (at points other than
$\sigma = 0$) is given by,

\begin{equation}
ds_{I}^2 = -\alpha^{2}
dn^{2}(\bar \sigma)[- d\tau^2 + d\sigma^2]
\end{equation}

for the solution inside the horizon. The roles of $\tau$ and 
$\sigma$ inside the horizon are reversed
in a way similar to that of $r$ and $t$ for the background metric.
The worldsheet is everywhere non--singular in this case except for
the  $\delta$ function in the Riemann curvature at $\sigma =0$ ($r=0$).

One can construct a string solution which
would reside entirely outside the horizon of the black hole.
This turns out to be given as :

\begin{equation}
r(\sigma) = \frac{l}{k}\frac{dn(\bar\sigma)}{cn(\bar\sigma)}
\end{equation}
with  $\bar\sigma = \pm\frac{
\alpha(\sigma -\sigma_{0})}{l}$. 
    
The embedding function is plotted in 
Fig. 2. It can be interpreted as a semi--infinite string with its
only boundary at $\sigma =0$ ($r=r_{1}$). The boundary condition
is obviously satisfied at this edge which corresponds to a
minimum of the embedding function (i.e. $r'(\sigma = 0) = 0$. Note
the curious fact that the boundary is not exactly at the horizon
(which correponds to $r=l=1$ in Figure 2) but slightly above it
(more precisely, at the value $r=\frac{l}{k}$).

The induced metric for this worldsheet is :
 
\begin{equation}
ds_{I}^{2} = \beta^{2}l^{2} \frac{sn^{2}(\bar \sigma)}
{cn^{2}(\bar \sigma)} [-d\tau^{2} + d\sigma^{2}]
\end{equation}

It is easy to see from the evaluation of the Ricci scalar that the 
worldsheet becomes singular at the only boundary of this semi--infinite
string. 

\section{\bf Perturbations}

\subsection{\bf  Perturbations for planetoids in general spherically 
symmetric, static backgrounds}

We now consider perturbative deformations of world-sheet 
in the manifestly covariant 
formalism {\cite{glfc:ref}}. 
The background metric is the usual, spherically symmetric,
static one mentioned in section II. We have two unknown
functions $b(r)$ and $\psi(r)$, which we specify according to our 
choice of the background.

The equations governing the perturbations are the Jacobi equations
and are related to the second variation of the Nambu--Goto action
evaluated at its stationary points.
A general perturbation of the embedding function $x^{\mu}(\sigma,\tau)$
can be written as $\delta x^{\mu} = E^{\mu}_{a}\phi^{a}
+n^{\mu}_{i}\phi^{i}$ where $\phi^{a}$ and $\phi^{(i)}$ are the perturbations
along the a--th tangent and the i-th normal respectively. We ignore the
tangential perturbations because they are essentially related to the
reparametrisation of the worldsheet and do not cause any deformation of
the worldsheet geometry, which is invariant under such transformations.
The equations satisfied by the quantities $\phi^{i}$ are given as 
{\cite{glfc:ref}}:

\begin{equation}
\Box \phi^{(i)} + (M^{2})^{i}_{j} \phi^{(j)} = 0
\end{equation}

However, it should be mentioned that in the most general setting, the
second variation of the Nambu-Goto action does involve quantities like
the components of the normal fundamental form 
$\mu_{ij}^{a}=g_{\mu\nu}n^{\mu}_{i}E^{\rho}_{a}D_{\rho}n^{\nu}_{j}
$. With the
choice of the embedding (planetoid)
 and the normals given below (for $\delta =0$
) it can be shown quite easily that all components of the normal
fundamental form are identically equal to zero irrespective of the
specific form of the functions $b(r)$ and $\psi(r)$ in the background
metric.  
Therefore we can use these equations for the $\phi^{(i)}$ in our
analysis of the perturbations.  

For strings in four dimensional backgrounds, these equations
constitute a pair of coupled differential equations,
with the quantity $(M^{2})^{ij}$ given as :

\begin{equation}
\left (M^{2}\right )^{ij} = K^{ab i}K_{ab}^{j} +
R_{\alpha\beta\mu\nu}E^{\alpha a}
n^{\beta i} E^{\mu}_{a} n^{\nu j} 
\end{equation}
with $K^{abi} = -g_{\mu\nu}\left ( E^{\alpha}_{a}D_{\alpha}E^{\mu}_{b}\right )
n^{\nu i} $, being the extrinsic curvature tensor of the worldsheet
along the i--th normal direction, $R_{\mu\nu\rho\sigma}$ the Riemann tensor
for the background spacetime, $ E^{\mu}_{a}$ the tangents in an
orthonormal frame and $n^{\mu i}$ the normals.  
$D_a$ is the world-sheet projection of the space-time covariant 
derivative $D_{\mu}$, where $D_a = E_a^{\mu} D_{\mu}$, 
$\mu$, $\nu = 0, 1, \cdots N -1$; $a = 1, 2, \cdots D
$, ($ a = \tau$, $\sigma$ for a string worldsheet ); $i = 1, 2, \cdots 
N -D$ (where $D$ is the number of world-sheet indices). We shall denote 
the second term in the R.H.S of the above expression
in future discussion as $A^{ij}$. 
Evaluation of the quantity $(M^{2})^{ij}$ thus 
depends on the general expressions for the tangents and
normals to the worldsheet. These are taken to be, 

\begin{equation}
E^{\mu}_{\tau} \equiv \left ( \frac{\alpha}{\sqrt{e^{2\psi}
\alpha^{2} - \beta^{2}
r^{2}}} , 0 , 0, \frac{\beta}{\sqrt{e^{2\psi}
\alpha^{2} - \beta^{2}r^{2}}}\right )
\quad ; \quad E^{\mu}_{\sigma} \equiv \left ( 0, \sqrt{1-\frac{b}{r}}
, 0, 0 \right )
\end{equation}

\begin{equation}
n^{\mu 1} \equiv \left ( \frac{r\beta e^{-\psi} \sin \delta}
{\sqrt{e^{2\psi}
\alpha^{2} - \beta ^{2}r^{2}}}, 0, \frac{1}{r}\cos \delta , 
\frac{\alpha e^{\psi}
\sin \delta}{r \sqrt{e^{2\psi}\alpha^{2} - \beta^{2}r^{2}}}
\right )
\end{equation}

\begin{equation}
n^{\mu 2} \equiv \left ( \frac{r\beta e^{-\psi} \cos \delta
}{\sqrt{e^{2\psi}\alpha ^{2} - \beta ^{2}r^{2}}},
0,-\frac{1}{r}\sin \delta , \frac{\alpha e^{\psi}\cos \delta}
{r \sqrt{e^{2\psi}
\alpha^{2} - \beta^{2}r^{2}}}
\right )
\end{equation}
where $\delta$ is an arbitrary angular parameter. For different values of 
$\delta$  we have
different normals. However all of them are related to each other by
$O(2)$ transformations. (In a general $N$ dimensional background with
a $D$ dimensional object living in it there is an $O(N-D)$ gauge
freedom in the choice of normals.) 
More specifically,

\begin{equation}
{\bar n}^{\mu i} = R^{ij} n^{\mu j}
\end{equation}

where $R^{ij}$ is the $O(N-D)$ dimensional rotation matrix (in our case
we have an $O(2)$ matrix).

We shall work with $\delta = 0$ and then write down all expressions
for a general $\delta$ by using its transformation properties.

The expressions for the $K_{ab}^{i}$
are given as follows. The only
nonzero components are the $K_{\tau \sigma}^{2}$ and $K_{\sigma \tau}^{2}$
which are ofcourse equal.

\begin{equation}
K_{\sigma \tau}^{2} = K_{\tau\sigma}^{2} =
 \frac{\alpha\beta e^{\psi}\sqrt{1-\frac{b}{r}}}{e^{2\psi}
\alpha^{2} - \beta^{2}r^{2}} \left ( r\psi^{\prime} - 1 \right )
\end{equation}

We also need to evaluate the quantity $ A^{ij} = 
R_{\alpha\beta\mu\nu}E^{\alpha a}
n^{\beta i} E^{\mu}_{a} n^{\nu j}$ 

As can be seen easily, the $A^{ij}$ for $i\neq j$ are all zero.
For $A^{11}$ and $A^{22}$ we have the following expressions :

\begin{equation}
A^{11} = \frac{b^{\prime} r - b}{2r^{3}} +  \frac{1}{r \left (e^{2\psi}
\alpha^{2} - \beta^{2}r^{2} \right )} \left [ e^{2\psi} \psi ^{\prime}
\left ( 1 - \frac{b}{r} \right ) \alpha^{2} - \beta^{2} b \right ]
\end{equation}

\begin{equation}
A^{22} = -\frac{\psi^{\prime}}{r} \left ( 1 -\frac{b}{r} \right )
+ \frac{e^{2\psi} \alpha^{2}}{e^{2\psi} \alpha^{2} - \beta ^{2} r^{2}}
\frac{b^{\prime}r - b}{2r^{3}} + \frac{r^{2}\beta^{2} \left (1-\frac{b}{r}
\right )}{e^{2\psi}\alpha^{2} - \beta^{2}r^{2}} \left [ \psi^{\prime\prime}
- \frac{b^{\prime}r - b}{2r(r-b)}\psi^{\prime} + {\psi^{\prime} }^{2} \right ]
\end{equation}

We now need to write down the full expression for the quantity for
$(M^{2})^{ij}$ This turns out to be ,

\begin{equation}
\left (M^{2}\right )^{ij} = K^{ab i}K_{ab}^{j} + A^{ij}
\end{equation}

Since $(M^{2})^{11}$ is equal to $A^{11}$ we write down the
expression for $(M^{2})^{22}$ only.

\begin{eqnarray}
\left ( M^{2} \right ) ^{22} =  
- 2\left ( \frac{\alpha\beta e^{\psi}\sqrt{1-\frac{b}{r}}}{e^{2\psi}
\alpha^{2} - \beta^{2}r^{2}} \left ( r\psi^{\prime} - 1 \right ) \right )^{2}
 -\frac{\psi^{\prime}}{r} \left ( 1 -\frac{b}{r} \right ) \nonumber \\
+ \frac{e^{2\psi} \alpha^{2}}{e^{2\psi} \alpha^{2} - \beta ^{2} r^{2}}
\frac{b^{\prime}r - b}{2r^{3}} + \frac{r^{2}\beta^{2} \left (1-\frac{b}{r}
\right )}{e^{2\psi}\alpha^{2} - \beta^{2}r^{2}} \left [ \psi^{\prime\prime}
- \frac{b^{\prime}r - b}{2r(r-b)}\psi^{\prime} + {\psi^{\prime} }^{2} \right ]
\end{eqnarray}

Let us now move on towards writing down the expressions for a
general $\delta$.
We shall denote quantities defined with respect to the new normal
with an overbar.
$K_{ab}^{i}K^{ab j}$ is denoted as $B^{ij}$. Therefore we have :

\begin{equation}
{\bar B}^{11} = \sin ^{2} \delta B^{22} \quad ; \quad
{\bar B}^{12} = \sin \delta \cos \delta B^{22} \quad ; \quad
{\bar B}^{22} = \cos ^{2} \delta B^{22}
\end{equation}

Similarly for $\bar A^{ij}$ we have :

\begin{eqnarray}
\bar A^{11} = \cos ^{2} \delta A^{11} + \sin ^{2} \delta A^{22} \\
\bar A^{22} = \sin ^{2} \delta A^{11} + \cos ^{2} \delta A^{22} \\
\bar A^{12} = \sin \delta \cos \delta \left ( A^{22} - A^{11} \right )
\end{eqnarray}

Now, by virtue of the presence of the off diagonal terms $A^{12}, B^{12}$
we will have genuinely coupled equations which govern the perturbations
of the planetoid solution. We shall however confine ourselves to
$\delta =0$ where the equations are uncoupled and easier to solve.
It must be admitted however, that a completely general
treatment of perturbations should be done with an
arbitrary $\delta$ with $\delta$ depending on $\sigma$ and $\tau$ as well.
In the latter case, the above equations, which correspond to rigid
rotations of the normal frame will naturally be modified.

{\sf (1) Minkowski spacetime in spherical coordinates}

We now analyse the perturbations about the planetoid string configuration
in spherically symmetric Minkowski spacetime. 
Recall that the string configuration (with $\sigma_{0} =0$)
 is given as :

\begin{equation}
t= t_{0} + \alpha \tau \quad ; \quad r= \frac{\alpha}{\beta} \vert \sin \beta
\sigma \vert 
\quad ; \quad \theta = \frac{\pi}{2} \quad ; \quad \phi = \phi_{0} +
\beta \tau
\end{equation}

We shall confine ourselves to the domain of $\sigma$ given as :
$0<\sigma<\frac{\pi}{2\beta}$ or $-\frac{\pi}{2\beta}<\sigma<0$.

The tangents $E^{\mu}_{a}$ and normals $n^{\mu i}$
to the worldsheet are, 

\begin{equation}
E^{\mu}_{\tau} \equiv \left ( \frac{\alpha}{\sqrt{\alpha^{2} - \beta^{2}
r^{2}}} , 0 , 0, \frac{\beta}{\sqrt{\alpha^{2} - \beta^{2}r^{2}}}\right )
\quad ; \quad E^{\mu}_{\sigma} \equiv \left ( 0, 1, 0, 0 \right )
\end{equation}

\begin{equation}
n^{\mu 1} \equiv \left ( 0, 0, \frac{1}{r} , 0 \right ) \quad ; \quad
n^{\mu 2} \equiv \left ( \frac{r\beta}{\sqrt{\alpha^{2} - \beta^{2}r^{2}}},
0, 0, \frac{\alpha}{r \sqrt{\alpha^{2} - \beta^{2}r^{2}}}
\right )
\end{equation}

Note that $(E^{\mu}_{a}, n^{\mu i})$ form an orthonormal
spacetime basis.

The nonzero $K_{ab}^{i}$ are given as :

\begin{equation}
K_{\sigma \tau}^{2} = K_{\tau\sigma}^{2} =
- \frac{\alpha\beta}{\alpha^{2} - \beta^{2}r^{2}}
\end{equation}

Note that the extrinsic curvature tensor components also diverge at the
edges of the string world--sheet. Therefore, if we evaluate the
mean curvature we will find indeterminate quantities appearing 
at the edge values of $\sigma$. The Nambu--Goto string worldsheet
is ill--defined at the edges.

Therefore we have,

\begin{equation}
\left ( M^{2} \right )^{22} = K^{\sigma\tau 2}K^{2}_{\sigma\tau}
+ K^{\tau\sigma 2}K^{2}_{\tau\sigma} = -\frac{2\beta^{2}}{\alpha^{2}
\cos^{4}{\beta\sigma}}
\end{equation}

The other entitities in the $(M^{2})^{ij}$ matrix are all zero.
Hence the perturbation equation for $\phi^{(2)}$ is given as
below :

\begin{equation}
-\frac{\partial^{2}{\phi^{(2)}}}{\partial\tau^{2}} + \frac{\partial^{2}
\phi^{(2)}}{\partial \sigma^{2}} - \frac{2\beta^{2}}{\cos^{2}\beta\sigma} 
\phi^{(2)} = 0
\end{equation}

On separating variables we have the harmonic oscillator equation
for the $\tau$ variable while the equation for the $\sigma$
variable turns out to be :

\begin{equation}
\frac{d^{2}\Sigma}{d\sigma^{2}} + \left (\omega^{2} - \frac{2\beta^{2}}{\cos
^{2}\beta\sigma} \right ) \Sigma = 0
\end{equation}

A simple pair of linearly independent solutions 
to the perturbation equation can be obtained
for $\omega = 0$. This is given as :

\begin{equation}
\Sigma = \tan \beta \sigma \qquad ; \qquad \Sigma = \beta\sigma \tan \beta
\sigma + 1
\end{equation}

Note that both these solutions have
 a divergence at $\sigma = \frac{\pi}{2\beta}$
This is due to the singular edges of the worldsheet geometry of the 
planetoid configuration. The
lowest mode of perturbation results in a solution which blows
up at the edges of the worldsheet. To obtain information about
the higher modes one needs to look at the general solution of the
relevant equation.
General solutions for eigenvalues ($\omega_{n}^{2}$)
are known to exist for the Schrodinger equation with a
potential $V(\sigma) = 2\beta^{2} sec^2 {\beta
\sigma}$ in quantum mechanics. There, this potential is known  
as the
Poschl--Teller I potential. The eigenvalues and eigenfunctions (unnormalised)
are {\cite{cks:pr}},

\begin{equation}
\omega^{2}_{n} = 4n^{2}\beta^{2}
\end{equation}

\begin{equation}
\Sigma_{n}(\beta;\sigma)  = \left ( \frac{1 - \gamma}{1 + \gamma}
\right )^{\frac{1}{2}} P_{n}^{\frac{1}{2},\frac{-3}{2}} (\gamma)
\end{equation}
where the $P_{n}$ denotes the Jacobi polynomial and $\gamma = 1 - 2\sin ^{2}
\beta \sigma$. 

The explicit expressions for the higher values of $n$ all
contain an overall factor $\tan \beta \sigma$. Therefore,
at $\sigma = \frac{\pi}{2\beta}$ there will always be
a divergence which is caused by the singularity in the
planetoid solution. If, however, as mentioned before we restrict 
ourselves to
the domain $-\frac{\pi}{2\beta} < \sigma < 0 $ or $0<\sigma <\frac{\pi}{2\beta}$
where the Nambu--Goto string is well defined in both the extrinsic
and intrinsic sense 
then, ofcourse, there is no problem with stability. 

The perturbation $\phi^{(1)}$ satisfies a simple
wave equation whose solutions are trivial. Note also the fact
that the generalised Raychaudhuri equation for the planetoid
strings would also be the same with the $\phi^{(2)}$ replaced 
by $F$ ($\theta_{a} = \frac{\partial_{a} F}{F}$ ).
Therefore, knowing the solutions of the $\Sigma$ equation would
imply solving both for perturbative as well as non--perturbative
deformations of the string configuration. 

{\sf (2) Spherical Rindler Spacetime}

The perturbation equations for the planetoid string in spherical
Rindler spacetime are given as (these are exclusively the 
equations for the $\sigma$ part of the perturbation $\phi ^{(i)} 
= T^{(i)}(\tau)\Sigma ^{(i)}(\sigma)$.
The $\tau$ part yields the usual harmonic oscillator equations.
The $\sigma$ equations are also harmonic oscillator equations
with different constants acting as the spring constant. These are,
 
\begin{equation}
\frac{d^{2}\Sigma^{(1)}}{d\sigma^{2}} + \left ( \omega^{2} + \alpha^{2}
\right ) \Sigma^{(1)} = 0
\end{equation}

\begin{equation}
\frac{d^{2}\Sigma^{(2)}}{d\sigma^{2}} + \left ( \omega^{2} - 
 \left ({\alpha^{2}-\beta^{2}} \right ) 
\right ) \Sigma^{(2)} = 0 
\end{equation}

The solutions to these equations are trivial. They are exponential 
or oscillatory according to the value of $\omega ^{2}$. For
the $\Sigma^{(1)}$, we just need $\omega ^{2} > 0$ for
oscillatory solutions. On the other hand, for
$\Sigma^{(2)}$ we need $\omega^{2} > (<)  \alpha^{2} - \beta^{2}$ 
for oscillatory (exponential) solutions.

\section{\bf Semi-classical Quantization}

 Semi-classical quantization of strings were performed in 
{\cite{vls:semi}} following 
the prescription given by Dashen {\it etal} {\cite{dhl:semi}}for 
time periodic solutions
in quantum mechanics and quantum field theory. In order to quantize 
the string solutions semi-classically, one needs to compute the classical
action of solutions $S_{cl}$ as a function of string mass $m$, where,
\begin{equation}
m = - \frac{dS_{cl}}{dT}
\end{equation}
and $T$ is the period in physical time given by, $T = \frac{2\pi\alpha}
{\beta}$. Hence, a knowledge of classical solutions is necessary for 
semi-classical quantization. Using (3), the 
expression for the classical action of solutions is given as,

\begin{equation}
S_{cl} = - \frac{2 T}{\pi\alpha'}\int_{r_{min}}^{r_{max}} dr g_{rr}
{\sqrt{- \tilde V(r)}}
\end{equation}
where, $r_{min}$ and $r_{max}$ are the minimum and maximum radius 
reached by the string respectively. The idea is to use the functional 
formulation of the WKB approximation via path integrals. The functional
integral is evaluated in a stationary phase approximation, where one
integrates over a function space and the stationary phase points are the
functions which satisfy the classical equation of motion and are periodic
solutions. The reduced action $W(m)$ is defined as, 
\begin{equation}
W \equiv S_{cl}(T(m)) + m T(m) = \frac{4}{T\alpha'}\int_{r_{min}}^{r_{max}}
dr \frac{T g_{t\phi} + 2\pi g_{\phi\phi}}{\sqrt{- \tilde V(r)}}
\end{equation}

The quantization condition is given by, $W = 2\pi n$. However, for the 
class of solutions being discussed here, the quantization condition 
is equivalent to 
\begin{equation}
W = 2\pi J
\end{equation}
where $J$ is the string angular momentum obtained by integrating 
the conserved world-sheet current $J_{\mu}$ ($\mu = \tau$, $\sigma$). 
The expression for $J$ is given by,

\begin{equation}
J = \frac{2}{\pi\alpha'}\int_{r_{min}}^{r_{max}} dr \frac{g_{t\phi}
+ \frac{2\pi}{T} g_{\phi\phi}}{\sqrt{- \tilde V(r)}}
\end{equation}

We now compute these quantities for the Ellis geometry, spherical
Rindler spacetime and the  $2 + 1$ 
dimensional black hole backgrounds to obtain the spectrum. We write 
down the quantization conditions in each of these cases and the relation
 between
mass and angular momentum turns out to lead to non-linear Regge 
trajectories ($\alpha' m^2 \neq 4 J$). In Minkowski space, one gets a 
linear Regge trajectory {\cite{dve:prd96}}

{\sf (1) Ellis Geometry}

For the case of Ellis geometry, the expressions for the physical 
quantities are given by,
\begin{equation}
S_{cl} = \frac{\pi}{\alpha^{\prime}} \left ( b_{0}^{2} -\frac{\alpha^{2}}
{\beta^{2}} \right ) 
\end{equation}
\begin{equation}
W = \frac{\pi}{\alpha^{\prime}} \left ( b_{0}^{2} +\frac{\alpha^{2}}{\beta^{2}}
\right )
\end{equation}
Therefore, the mass $m$ is given by,
\begin{equation}
m = \frac{\alpha}{\alpha^{\prime}\beta} 
\end{equation}

Evaluating $J= \frac{W}{2\pi}$ one gets the curious relation,
\begin{equation}
2J = \alpha^{\prime}m^{2}  + \frac{b_{0}^{2}}{\alpha^{\prime}}
\end{equation}

This is a linear Regge relation with an intercept  proportional
to the square of the throat radius of the wormhole.
It should be mentioned that by naively putting $b_{0} = 0$
in the Ellis geometry one does not get back Minkowski spacetime --
 but a couple
of Minkowski worlds connected by a throat which has gone singular
by the assumption $b_{0} = 0$. This is also reflected in the
Regge relation where we notice a missing factor of $2$ in
comparison with the Minkowski spacetime result $4J = \alpha^{\prime}m^{2}$.
  
{\sf (2) Spherical Rindler spacetime}

The expressions $S_{cl}$, reduced action $W(m)$, mass and string 
angular momentum are given by,

\begin{equation}
S_{cl} = - \frac{T}{\pi\alpha'} {\sqrt{1 - \frac{4\pi^2}{T^2}}};
\qquad W(m) = \frac{4\pi}{T\alpha'} \frac{1}{\sqrt{1 - \frac{4\pi^2}{T^2}}}
\end{equation}

And,
\begin{equation}
m = \frac{1}{\pi\alpha'{\sqrt{1 - \frac{4\pi^2}{T^2}}}}; \qquad
J = \frac{2}{T \alpha'{\sqrt{1 - \frac{4\pi^2}{T^2}}}}
\end{equation}
One can again note from the above relations that $\alpha' m^2$ is not
proportional to $J$. Infact, as can be easily seen, $J = \frac{\beta}
{\alpha} m $.

{\sf (3) 2 + 1 dimensional black hole space-time}

The physical potential for $2 + 1$ dimensional black hole (assuming
$M = 1, J =0$) is given by,
\begin{equation}
\tilde V(r) = \left ( 1 - \frac{r^2}{l^2} \right )
\left [\frac{r^2}{l^2} - 1 - \frac{4\pi^2 r^2}{T^2}
\right ]
\end{equation}

In this case, there are two kinds of planetoid strings, one which starts 
from 
$r = 0$ and ends precisely at the horizon having a finite length. 
The other one which starts at a point away from the horizon and extends
upto infinity with an infinite length. We shall consider the planetoid 
strings which are inside the horizon. For these strings, the maximum
radius $r_{max} = l$ and $r_{min} = 0$. 
 The classical action of the solution is, 
\begin{equation}
S_{cl} = \frac{2 T l}{\pi\alpha'} E(k) 
\end{equation}
where $E(k)$ is the complete elliptic integral and $k = {\sqrt{1 - 
\frac{4\pi^2 l^2}
{T^2}}}$ is the elliptic modulus {\cite{gr:tab}}. $W$ is given by the expression, 
\begin{equation}
W = \frac{8\pi l^3}{T\alpha' k^2}\left [ K(k) - E(k)\right ]
\end{equation}
We can now compute the mass $m$ and angular momentum $J$ from the above 
expressions and they are given by, 
\begin{equation}
m = \frac{l}{\pi T^{2}\alpha^{\prime}k^{2}} \left [ 
- 2T^{2} E(k) + 8\pi^{2}l^{2} K(k) \right ]
\end{equation}

\begin{equation}
J = \frac{4 l^3}{T\alpha' k^2}\left [ K(k) - E(k)\right ]
\end{equation}
Since $J$ is not proportional to $m^2$ it leads to a nonlinear Regge 
trajectory. For $T \sim 2\pi l$, the quantization condition 
reads $T^2 \sim 4\pi^2 n\alpha'$. In the above limit, $k \rightarrow
0$ and the mass of the string becomes $\alpha' m^2 \simeq 4 n$, where 
one recovers the linear behaviour of the Regge trajectory. 

The invariant string length is given by,

\begin{equation}
s = 2 \int_0^{r_{max}} dr{\sqrt{g_{rr}}} = \pi l
\end{equation}

On the other hand, if we consider the other planetoid string where 
$r_{min} = \frac{l^2}{1 - \frac{4\pi^2 l^2}{T^2}}$, then the string 
length becomes infinite as it stretches outside the horizon. 

\section{\bf Summary and Conclusions}

Let us first summarise the results we have obtained. Firstly,
we have new examples of planetoids in a variety of
of spacetime backgrounds. These include the Ellis geometry,
the spherical Rindler geometry and the spacetime of the $2+1$
dimensional BTZ black hole. A curious feature about most of
these planetoid string configurations is the presence of
a {\em worldsheet singularity} at the edges.  Thereafter, we write down the 
general equations governing perturbations of any planetoid
solution in a static, spherically background. The general
formalism is then applied to planetoids in Minkowski and
spherical Rindler spacetimes, where the planetoid perturbation
equations turn out to be exactly solvable. We comment on the
stability of these solutions by looking at divergences in the
solutions governing perturbations. The Minkowski spacetime
planetoid solution turns out to be stable in the nonsingular
domain of the worldsheet geometry where we can apply the
perturbation theory. The planetoid in 
spherical Rindler spacetime is also stable. We have not been able to
integrate the perturbation equation for the planetoids in
a $2+1$ dimensional black hole or the Ellis geometry essentially due to the
complicated form of the string configurations.  
Finally, using methods of semiclassical quantisation we
proceed to quantise the planetoids obtained in Section III.
The Regge relations are written down and the semiclassical
quantisation conditions are derived for each of these planetoids.
It turns out that in the Ellis geometry, the Regge relation
is linear but we have an intercept proportional to the
square of the wormhole throat radius. In spherical Rindler
spacetime $J$ is linearly related to $m$ while in the
BTZ black hole we cannot write down an explicit general relation
between $J$ and $m$.  

Obtaining further planetoid solutions would be obviously of
interest in future. A generalisation of the ansatz which could
be applicable in time-dependent backgrounds would also be
worth attempting (recall that in Section I, we showed that
the planetoid ansatz is incompatible with time--dependant
metrics of a certain general type).
The recent work on rigidly rotating strings {\cite{fhdv:hepth}}
proposes a more general class of strings of which planetoids
are a special case. It would be worth trying out new examples
of such rigidly rotating strings in curved backgrounds and
analyse the stability of these configurations. Finally, the
question of non--perturbative deformations which are governed
by the generalised Raychaudhuri equations must be addressed
in order to study the formation of cusps and kinks on the string
worldsheet. Planetoids and rigidly rotating strings could be
prototypes where these equations are solvable and information
can be obtained in atleast a specific context. 

\section*{\bf Acknowledgements}
S.K. acknowledges financial support from the Inter--University
Centre for Astronomy and Astrophysics, Pune, India. Work of S.M. is 
supported by Alexander von Humboldt foundation. The authors thank 
Norma Sanchez for pointing out ref.[7] to them and Riccardo
Capovilla for discussions.


\centerline{\bf FIGURE CAPTIONS}

{\bf Fig. 1} : The embedding function $r(\sigma)$ versus $\sigma$ for the
planetoid string inside the horizon of the BTZ black hole. The values for
the various parameters are : $\alpha = 1$, $\beta =.5$, $l=1$, $k^{2}=.75$   

{\bf Fig. 2} : The embedding function $r(\sigma)$ versus $\sigma$ for the
planetoid string outside the horizon of the BTZ black hole. The values for
the various parameters are : $\alpha = 1$, $\beta =.5$, $l=1$, $k^{2}=.75$   

\end{document}